\documentclass[%
 reprint,
 superscriptaddress,
 amsmath,amssymb,
 aps,
showkeys,
]{revtex4-2} 

\usepackage{graphicx}% Include figure files
\usepackage{dcolumn}% Align table columns on decimal point
\usepackage{bm}% bold math
\usepackage{xcolor}

\usepackage{braket}
\usepackage{bm}% bold math
\usepackage{amsmath} % allows you to display text inside the math-environment
\usepackage{xspace}	% allows you to put or not the space at the end of a newcomand
\usepackage{amsfonts}
\usepackage{comment}
\usepackage{tabularx}
\usepackage{textcomp}

\newcommand{\diff}{\mathop{}\!\mathrm{d}}

\newcommand{\derivd}[3]{\frac{\diff^{#3} #1}{\diff #2^{#3}}}
\newcommand{\CCW}{_{\scriptscriptstyle{\rm CCW}}}
\newcommand{\CW}{_{\scriptscriptstyle{\rm CW}}}
\newcommand{\CWCCW}{_{\scriptscriptstyle{\rm CW/CCW}}}
\newcommand{\dogG}{\left(-i\Delta\omega+\gamma+\Gamma\right)}

\newcommand{\Real}{\Re}
\newcommand{\Imag}{\Im}
\newcommand\nm{~{\rm nm}}
\newcommand\um{~\mu{\rm m}}%{~{\rm \mu m}}

\newcommand{\dB}{~{\rm dB}}

\newcommand{\GHz}{~{\rm GHz}}
\newcommand{\THz}{~{\rm THz}}

\newcommand{\mWHz}{~{\rm mW\cdot Hz}}
\newcommand{\kOhm}{~{\rm k}\Omega}%{~{\rm k\Omega}}

\newcommand{\figname}{Fig.~}
\newcommand{\figsname}{Figs.~}

\newcommand{\fignametot}{Figure~}
\newcommand{\figsnametot}{Figures~}
\newcommand{\appname}{Appendix~}

\newcommand{\eqname}{Eq.~}
\newcommand{\eqsname}{Eqs.~}
\begin{document}

%\preprint{APS/123-QED}

\title{An interferometric method to estimate the eigenvalues of a Non-Hermitian two-level optical system}

\author{Stefano Biasi $^\dagger$}
    \email{Corresponding author: stefano.biasi@unitn.it}
    \affiliation{Nanoscience Laboratory, Department of Physics, University of Trento, Via Sommarive 14, Povo - Trento, 38123, Italy}
    \altaffiliation{These authors contributed equally to this work.}%Lines break automatically or can be forced with \\
\author{Riccardo Franchi $^\dagger$}%
    \affiliation{Nanoscience Laboratory, Department of Physics, University of Trento, Via Sommarive 14, Povo - Trento, 38123, Italy}
   % \altaffiliation{These authors contributed equally to this work.}
\author{Filippo Mione}
 \affiliation{Nanoscience Laboratory, Department of Physics, University of Trento, Via Sommarive 14, Povo - Trento, 38123, Italy}
\author{Lorenzo Pavesi}
    \affiliation{Nanoscience Laboratory, Department of Physics, University of Trento, Via Sommarive 14, Povo - Trento, 38123, Italy}

\date{\today}% It is always \today, today,

\begin{abstract}
	Non-Hermitian physics has found a fertile ground in optics. Recently, the study of mode degeneracies, i.e. exceptional points, has led to the discovery of intriguing and counterintuitive phenomena. Degeneracies are typically modeled through the coupled mode theory to determine the behaviour of eigenstates and eigenvalues. However, the complex nature of the eigenvalues makes hard their direct characterization from the response spectrum. Here, we demonstrate that a coherent interferometric excitation allows estimating both the real and imaginary parts of the eigenvalues. We studied the clockwise and counter-clockwise modes in an optical microresonators both in the case of Hermitian and non-Hermitian intermodal coupling. We show the conditions by which a resonant doublet, due to the dissipative coupling of counter-propagating modes caused by surface roughness backscattering, merges to a single Lorentzian. This permits to estimate the optimal quality factor of the microresonator in the absence of modal coupling caused by backscattering. Furthermore, we demonstrate that a taiji microresonator working at an exceptional point shows a degeneracy splitting only in one propagation direction and not in the other. This follows from the strongly non-Hermitian intermodal coupling caused by the inner S-shaped waveguide.
\end{abstract}

%Use showkeys class option if keyword

%\begin{document}

\maketitle

\section{Introduction}
\label{sec:Intro}

In an open system, the energy can be exchanged with the environment. Its description by a Hermitian Hamiltonian is approximated, because the conservation of energy is not a-priori satisfied \cite{Non-HermitianPTsymmetry_Demetrios}. In this case, a non-Hermitian Hamiltonian has to be used which leads to many fascinating new phenomena \cite{Non-Hermitian_Ashida}. A radical difference between non-Hermitian (dissipative) and Hermitian (conservative) physics arises in the presence of degeneracies, i.e. when the Hamiltonian eigenvalues coalesce \cite{Degeneracy_Berry}. In the conservative case, these degeneracies are called \textit{diabolic points}, while in the dissipative one they are called \textit{exceptional points}. The former are characterized by real eigenvalues that coalesce while maintaining orthogonal eigenvectors \cite{Berry_Diabolic}. Differently, the latter are characterized by complex eigenvalues that coalesce simultaneously with the eigenvectors \cite{Non-Hermitian_Ashida}. Close to these exceptional points, counterintuitive and intriguing phenomena occur \cite{Degeneracy_Berry,ParytyTimeSymExcPoint_Lan}. Since the propagation of light is generally affected by losses, optics constitutes one of the most fertile grounds to study open systems \cite{Exceptional_Alu}. In recent years, the realization of optical structures working near exceptional points has allowed the demonstration of non-trivial effects such as unidirectional invisibility \cite{Large-scaleExceptionalpoint_Zhang, Taiji_Allegra, Invisibility_Demetrios}, loss-induced transparency \cite{Trasparency_Christodoulides}, directional emission \cite{ChiralModeLasing_Lan, Lasing_Albert} and the PT-symmetric laser \cite{MicroringLaser_Mercedeh}. Typically, these optical systems are based on microring resonators which are described within the temporal coupled mode theory \cite{ChiralModeLasing_Lan, Exceptional_Lan}. In these systems, the knowledge of the eigenvalues represents a fundamental aspect. However, their complex nature makes difficult a simple and direct estimate from their optical response spectra.

\begin{figure}[t!]
	\centering
	\includegraphics[scale=1]{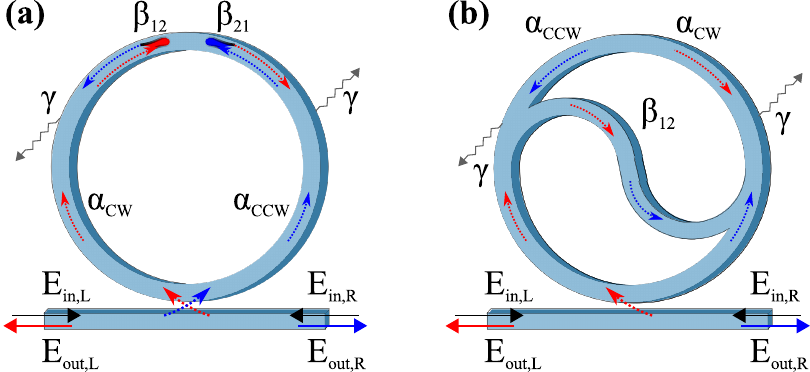}
	\caption{Sketch of a ring and a taiji microresonator coupled to a bus waveguide, (a) and (b) respectively. The arrows indicate the propagating fields. Specifically, the red (blue) highlights propagation in the clockwise (counter-clockwise) directions. The different symbols are defined in the text.}
	\label{fig:microresonators}
\end{figure}

In this work, we describe a method to estimate non-Hermitian eigenvalues based on a coherent interferometric technique to excite the modes of an optical microresonator. The coherent interferometric excitation technique is based on the simultaneous excitation of the clockwise and counter-clockwise modes of a microresonator coupled to a bus waveguide. Its key aspect is that the microresonator response depends on the relative phase difference of the signals input at the two sides of the bus waveguide (\figname\ref{fig:microresonators}). We applied the coherent interferometric excitation to representative Hermitian and non-Hermitian systems which are based on a simple microring or a taiji microresonator \cite{Taiji_Allegra}. In the former case, the coupling between the counter-propagating modes is given by the surface-wall roughness, i.e. the backscattering \cite{Back_Bogaerts}, while in the second, it is related to a S-shaped waveguide embedded into the taiji microresonator \cite{Taiji_Albert, taiji_Mercedeh}. 

In section \ref{sec:Theory}, we model the bus waveguide/microresonator system response as a function of the relative phase between the signals input in the forward and reverse directions. We show the merging of the resonant doublet due to the microresonator modes into a single Lorentzian. We propose an analytical expression for the eigenstates and eigenvectors. We demonstrate that, in both the Hermitian and non-Hermitian case, the coherent interferometric excitation can be used to determine the microresonator Q-factor in absence of the surface roughness losses. For Hermitian systems, we show that the coupling coefficients can be estimated from the spectral position of the single Lorentzian resonances.
For non-Hermitian systems, we demonstrate that the complex nature of the eigenvalues and their characteristic parameters can be studied by changing the relative intensities and phases of the input signals in the coherent interferometric excitation. In this framework, we show that the taiji microresonator works on an exceptional point and we show a direction dependent response due to the coeherent interferometric excitation. A resonance splitting is predicted only for transmission in the propagation direction which couples to the reflection of the inner S-shaped waveguide.

In section \ref{sec:Exp}, we implement the coherent interferometric techinque in a simple set-up and we study two different types of microresonators where non-Hermitian physics can be demontrated. We extract the different parameters (intrinsic and extrinsic loss coefficients, real and imaginary intermodal coupling coeficients) for a simple microring resonator and a taiji microresonator and we validate the model by fitting their transmission spectra.

\begin{figure*}[t!]
	\centering
	\includegraphics[width=1\textwidth]{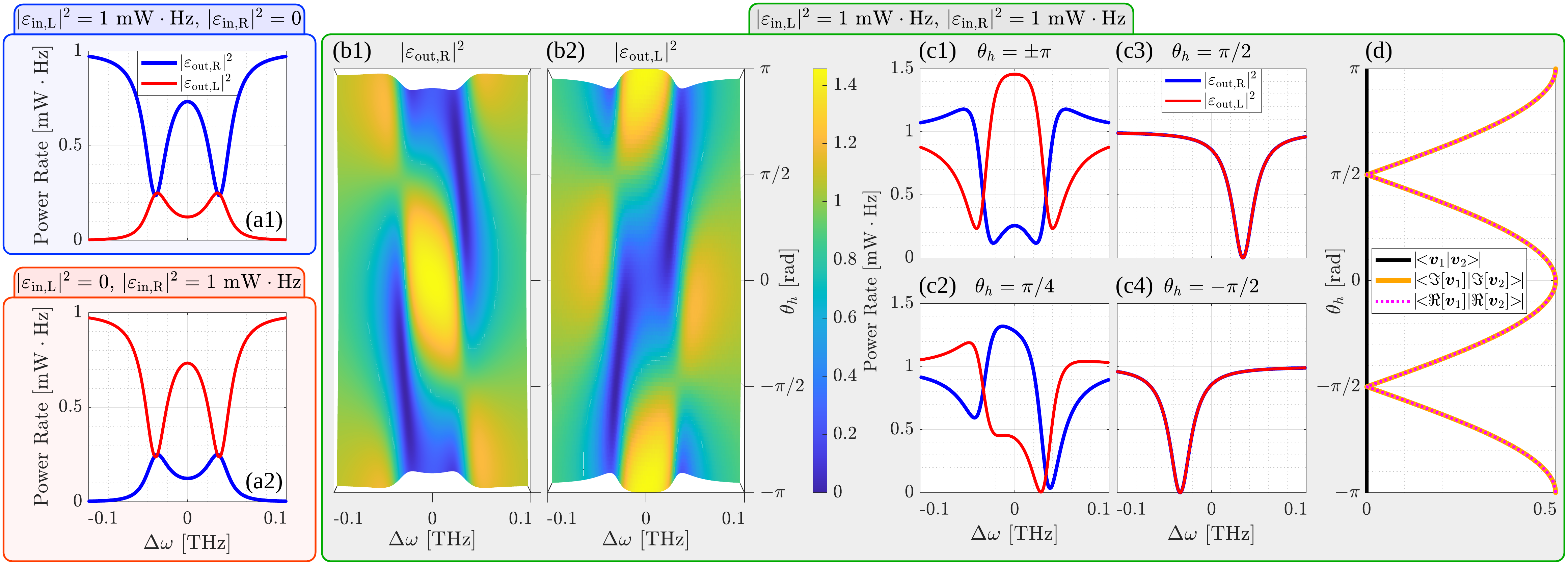}
	\caption{Theoretical results for an Hermitian coupling. Panels (a1)-(a2) show the transmission and reflection spectra for an excitation from the left and right input, respectively. Maps (b1) and (b2) show, respectively, the right and left output field intensities as a function of the angle ($\theta_{h}$) and of the detuning frequency ($\Delta \omega$) for a symmetric interferometric excitation. Panels (c1)-(c2) show the output field spectral lineshapes for fixed values of $\theta_{h}=\{\pm \pi,\pi/4,\pi/2,-\pi/2\}$. The red (blue) lines denote the left (right) output fields. The graph (d) shows the modulus of the inner product of the eigenstates and their real and imaginary parts as a function of $\theta_{h}$. The solid-black, solid-orange, and dashed-magenta lines refer to $|\braket{\bm{v}_1||\bm{v}_2}|$, $|\braket{\Imag[\bm{v}_1]|\Imag[\bm{v}_2]}|$, and $|\braket{\Real[\bm{v}_1]|\Real[\bm{v}_2]}|$. The coupling is Hermitian, and therefore, the two eigenstates are orthogonal, i.e. $\braket{\bm{v}_1|\bm{v}_2}=0$.  Here, we used the following coefficients: $\Gamma=\gamma=6.8\GHz$ and $\beta_{12}=-\beta_{21}^{*}=33.2\GHz$.% and $\omega_{0}=2\pi \cdot 193$ THz.
	}
	\label{fig:Plot_tot_Hermitian_th}
\end{figure*}

\section{Theoretical model}
\label{sec:Theory}

Let us consider the electric field amplitudes of the clockwise ($\alpha\CW$) and the counter-clockwise ($\alpha\CCW$) propagating modes in a simple or in a taiji microresonator coupled to a bus waveguide (\figname\ref{fig:microresonators} (a) and (b)). We model these systems with the temporal coupled mode theory and the properties of a time-reversal-invariant system \cite{TCMT_Deriv,TCMT_Fan}. Therefore, $\alpha\CW$ and $\alpha\CCW$ are solutions of \cite{Back_Biasi}: 
\begin{equation}\label{eq:TCMT}
\begin{split}
i \derivd{}{t}{}
\begin{pmatrix}
\alpha\CCW \\
\alpha\CW
\end{pmatrix}
= 
&\begin{pmatrix}
\omega_0-i(\gamma+\Gamma) & -i\beta_{12} \\
-i\beta_{21} & \omega_0-i(\gamma+\Gamma)
\end{pmatrix}
\begin{pmatrix}
\alpha\CCW \\
\alpha\CW
\end{pmatrix}\\
&- \sqrt{2\Gamma}
\begin{pmatrix}
E_{\rm in, L} \\
E_{\rm in, R}
\end{pmatrix}\,,
\end{split}
\end{equation}
where $\omega_0$ is the angular frequency of the microresonator resonance, $E_{\rm in,L}$ ($E_{\rm in,R}$) is the electric field amplitude input into the left (right) edge of the bus waveguide, $\gamma$ and $\Gamma$ are the intrinsic and extrinsic damping rates, $\beta_{12}$ and $\beta_{21}$ are the intermodal coupling coefficients. Precisely, $\gamma$ describes the losses of the cavity related to intrinsic factors, such as material absorption and bending, while $\Gamma$ refers to the coupling with the bus waveguide. It is worth noting that $\gamma$ and $\Gamma$ are real numbers while $\beta_{12}$ and $\beta_{21}$ are complex numbers. These latter describe the energy exchange between the counter-propagating modes, which can have either Hermitian (conservative) or non-Hermitian (dissipative) nature. In a simple microring, the intermodal energy exchange is caused by the backscattering, which yields complex coupling coefficients with independent real and imaginary parts \cite{Back_Biasi}. In a taiji microresonator, neglecting the backscattering, $\beta_{21}$ reduces to zero, while $\beta_{12}$ is defined by the inner S-shaped waveguide \cite{Taiji_Albert,Taiji_Allegra,Taiji_Franchi}. 

In the interferometric excitation technique, electric fields $E_{\rm in, L}:= \varepsilon_{\rm in,L} e^{- i\omega t}$ and $E_{\rm in, R}:= \varepsilon_{\rm in,R} e^{i\phi} e^{- i\omega t}$ are coherently input from both sides (left L, right R) of the bus waveguide and the transmitted fields $E_{\rm out, L}$ and $E_{\rm out, R}$ are measured to probe the system response. Note the phase difference $\phi$ between $E_{\rm in, R}$ and $E_{\rm in, L}$. The input and output fields are related by \cite{Back_Biasi,TCMT_Deriv}: 
\begin{equation}\label{eq:Eout}
\begin{pmatrix}
E_{\rm out, R} \\
E_{\rm out, L}
\end{pmatrix}
= 
\begin{pmatrix}
E_{\rm in, L} \\
E_{\rm in, R}
\end{pmatrix}
+i\sqrt{2\Gamma}
\begin{pmatrix}
\alpha\CCW \\
\alpha\CW
\end{pmatrix}\,,
\end{equation}
where in the steady-state $\alpha\CWCCW := a\CWCCW e^{- i\omega t}$. \eqsname\eqref{eq:TCMT} and \eqref{eq:Eout} lead to the following steady-state output field amplitudes:
\begin{align}\label{eq:tR}
\varepsilon_{\rm out, R} = &\left(1-\frac{2\Gamma\dogG}{\dogG^2-\beta_{12}\beta_{21}}\right)\varepsilon_{in,L}\\\nonumber
&+\left(\frac{2\Gamma\beta_{12}}{\dogG^2-\beta_{12}\beta_{12}}\right)e^{i\phi}\varepsilon_{\rm in,R}\\\label{eq:tL}
\varepsilon_{\rm out, L} = &\left(1-\frac{2\Gamma\dogG}{\dogG^2-\beta_{12}\beta_{21}}\right)e^{i\phi}\varepsilon_{\rm in,R}\\\nonumber
&+ \left(\frac{2\Gamma\beta_{21}}{\dogG^2-\beta_{12}\beta_{21}}\right)\varepsilon_{\rm in,L}\,,
\end{align}
where $\Delta\omega:=\omega-\omega_0$. \eqname\eqref{eq:tR} and \eqname\eqref{eq:tL} show that the output fields result from the sum of the transmitted and reflected input fields from the microresonator. Note the relevant role played by the phase term $\phi$ which allows tuning the excitation of the two microresonator modes. This role is better exemplified by introducing the super-modes:
$b_{1,2} := (\alpha\CCW \pm \alpha\CW e^{-i\phi})/\sqrt{2}$. Here, the sum and difference of the counter-propagating amplitudes are governed by $\phi$. In stationary condition, \eqname\eqref{eq:TCMT} can be expressed in the super-mode description as:
\begin{equation}\label{eq:TCMT_SuperModes}
\begin{split}
\omega
\begin{pmatrix}
b_{1} \\
b_{2}
\end{pmatrix}
= 
&\begin{pmatrix}
\omega_0-i(\gamma+\Gamma+g_{1}) & i g_{2} \\
-i g_{2} & \omega_0-i(\gamma+\Gamma-g_{1})
\end{pmatrix}
\begin{pmatrix}
b_{1} \\
b_{2}
\end{pmatrix}\\
&-\sqrt{2\Gamma}
\begin{pmatrix}
\frac{\varepsilon_{\rm in, L}+\varepsilon_{\rm in, R}}{\sqrt{2}} \\
\frac{\varepsilon_{\rm in, L}-\varepsilon_{\rm in, R}}{\sqrt{2}}
\end{pmatrix}\,,
\end{split}
\end{equation}
where $g_{1,2} := (\beta_{12}e^{i\phi}\pm\beta_{21}e^{-i\phi})/2$. Within the super-mode description, the microresonator excitation results from a linear combination of only the left and right input field amplitudes.
On the contrary, the coupling terms ($g_{1,2}$) depend on $\phi$, on $\beta_{12}$ and on $\beta_{21}$. \eqname\eqref{eq:TCMT_SuperModes} has eigenvalues: 
\begin{equation}\label{eq:eigenvalues}
\lambda_{1,2} = \omega_0\pm\sqrt{-\beta_{12}\beta_{21}}-i(\gamma+\Gamma),
\end{equation}
which do not depent on $\phi$ since they are an intrinsic property of the system. In contrast, the normalized eigenstates of \eqname\eqref{eq:TCMT_SuperModes}
\begin{equation}\label{eq:Eigenvectors_g}
v_{1,2} = \frac{1}{\sqrt{1+\left|\frac{g_{1}\mp\sqrt{g_{1}^2-g_{2}^2}}{g_{2}}\right|^2}}
\begin{pmatrix}
\frac{g_{1}\mp\sqrt{g_{1}^2-g_{2}^2}}{g_{2}}\\
1
\end{pmatrix}
\end{equation}
are affected by the interferometric excitation, i.e. $\phi$.

It is customary to define the Hermitian ($h$) and non-Hermitian ($n$) intermodal coupling coefficients \cite{Back_Biasi, Lasing_Albert} 
\begin{equation}\label{eq:h-n}
h := i\frac{\beta_{12}-\beta_{21}^{*}}{2}\,,\quad
n := \frac{\beta_{12}+\beta_{21}^{*}}{2}.
\end{equation}
These definitions allow us to discuss in the following Hermitian and non-Hermitian systems, separately.

\subsection{Hermitian coupling and diabolic point}

Here, $\beta_{12}=-\beta_{21}^*=\beta$, and therefore $h \neq 0$ and $n=0$. This generates a continuous conservative exchange of energy between the clockwise and the counter-clockwise modes inducing the typical splitting (doublet) in the transmission spectra  of a bus waveguide/microring system \cite{Back_Biasi}. \fignametot\ref{fig:Plot_tot_Hermitian_th} (a1)-(a2) shows the computed output spectral intensities for a real coupling coefficient $\beta=33.2\GHz$ when $\varepsilon_{\rm in,L}=1\,\sqrt{\mWHz}$ and $\varepsilon_{\rm in,R}=0$ (top spectra) or when $\varepsilon_{\rm in,L}=0$ and $\varepsilon_{\rm in,R}=1\,\sqrt{\mWHz}$ (bottom spectra), i.e. in a standard, single excitation side, experiment. Hence, in the top graph, $|\varepsilon_{\rm out,R}|^2$ and $|\varepsilon_{\rm out,L}|^2$ are the transmitted and reflected intensities while the reverse holds for the bottom graph. The output fields are not simple Lorentzians, but show a balanced doublet (same peak intensities) due to the interaction between the counter-propagating modes in the microring \cite{BalancedSplitting_Li}. Lorentz reciprocity theorem \cite{Reciprocity_Potton, Isol_Renner} assures equal transmission spectra and the Hermitian coupling yields equal reflection spectra for both input (L or R) directions. 

The situation changes when the interferometric excitation is used. In the Hermitian case, as shown in \appname\ref{app:Hermitian}, \eqname\eqref{eq:Eigenvectors_g} can be reformulated as:
\begin{equation}\label{eq:Eigenvectors_h}
v_{1,2} = \frac{1}{\sqrt{1+\left|\frac{\sin{\left(\theta_h\right)}\mp1}{\cos{\left(\theta_h\right)}}\right|^2}}
\begin{pmatrix}
i \frac{\sin{\left(\theta_h\right)}\mp1}{\cos{\left(\theta_h\right)}}\\
1
\end{pmatrix}\,,
\end{equation}
where the angle $\theta_h=\phi+arg[\beta]$ is introduced to show that the eigenstates do not depend on the strength of the coupling ($|\beta|$) but only on its phase. As expected for an Hermitian system, its eigenvectors are orthogonal and, therefore, their inner product $|\braket{ \bm{v}_1|\bm{v}_2}| = 0$ for any $\phi$. On the other hand, its eigenvalues depend on $|\beta|$:
\begin{equation}\label{eq:omega12_beta}
\lambda_{1,2} = \omega_{0}\pm |\beta|-i(\gamma+\Gamma),
\end{equation}
but not on $\phi$. Their imaginary parts ($\Imag[\lambda_{1,2}]= -\gamma - \Gamma$) are related to the losses, while their real parts ($\Real[\lambda_{1,2}]=\omega_{1,2}=\omega_0 \pm |\beta|$) describes the splitting of the resonance as a result of the coupling between the counter-propagating (back-reflected) modes. When $|\beta|=0$, the system works on a diabolic point \cite{Degeneracy_Berry}.
Indeed, the eigenvalues are degenerate ($\omega_{1}=\omega_{2}=\omega_{0}$), the eigenvectors are orthogonal and the resonance splitting is linear with $|\beta|$. 

Using a symmetric (i.e. same field intensities) interferometric excitation ($\varepsilon_{\rm in,L} = \varepsilon_{\rm in,R} = 1 \sqrt{\mWHz}$), $\varepsilon_{\rm out,L}$ and $\varepsilon_{\rm out,R}$ depend on $\theta_{h}$. Panels (b1) and (b2) of \figname\ref{fig:Plot_tot_Hermitian_th} show the transmitted field intensities as a function of $\Delta \omega$ and $\theta_h$. A rich dynamics is observed with transmitted intensities which increase or decrease depending on the excitation conditions. Note that since the sum of the input intensities is equal to 2 ${\mWHz}$ it is not suprising to find transmitted intensities larger than 1 ${\mWHz}$. \figsnametot\ref{fig:Plot_tot_Hermitian_th} (c1)-(c4) show the spectra of the transmitted intensities for $\theta_{h} = \{\pm\pi,\pi/4,\pi/2,-\pi/2\}$. As a function of $\theta_{h}$, different transmission lineshapes at the bus waveguide outputs are observed. 

\figname\ref{fig:Plot_tot_Hermitian_th} (d) shows the modulus of the inner product of the eigenstates ($|\braket{\bm{v}_1|\bm{v}_2}|$) and their real ($|\braket{\Real[\bm{v}_1]| \Real[\bm{v}_2]}|$) and imaginary ($|\braket{\Imag[\bm{v}_1]|\Imag[\bm{v}_2]}|$) parts  as a function of $\theta_h$. Specifically, the inner product of  the real and imaginary components follow a periodic function proportional to $|\cos{[\theta_h]}|/2$, changing their mutual projection as $\theta_{h}$ varies. When the real and imaginary parts of the eigenstates are parallel, i.e. $\theta_{h} = \{-\pi, 0, \pi\}$, the transmission doublet is balanced (see \figname\ref{fig:Plot_tot_Hermitian_th} (c1)), and there is a symmetric energy exchange between the counter-propagating modes. On the contrary, when $\theta_{h}=\pm\pi/2$, the transmission doublet merges into a single resonance and the same spectra are observed from both waveguide ends. In these peculiar cases, both the real and imaginary parts of the eigenstates become orthogonal and their inner products reduce to zero. For these $\theta_{h}$ values, both the left and right output intensities take the shape of a single Lorentzian.

Under these conditions, i.e. for $\theta_{h}=\phi+arg[\beta]=\pm\pi/2+2\pi m$ ($m\in \mathbb{Z}$) and $\varepsilon_{0} = \varepsilon_{\rm in,R} = \varepsilon_{\rm in,L}$, \eqsname\eqref{eq:tR} and \eqref{eq:tL} reduce to the typical analytical expression of a Lorentzian \cite{Time_Biasi, Optical_Heebner}:
\begin{equation}\label{eq:Lor}
|\varepsilon_{\rm out,R}|^2 = |\varepsilon_{\rm out,L}|^2= 
\left(1 - \frac{4 \gamma \Gamma}{(\Delta \omega \mp |\beta|)^2+(\gamma + \Gamma)^2} \right)
|\varepsilon_{0}|^2\,.
\end{equation}

Let us discuss this result. \eqsname\eqref{eq:Lor} shows that the transmission of the bus waveguide/microresonator system is the same as the one obtained in the absence of the intermodal coupling (i.e. no backscattering). The position of the Lorentzian is spectrally shifted  by the intermodal coupling strength with respect to the microresonator resonance. Thus, the interferometric excitation permits to extract relevant parameters directly from the measured transmission spectra:
\begin{enumerate}
	\item the amplitude $|\beta|$, by measuring the resonant frequencies of the Lorentzians;
	\item the phase $arg[\beta]$, by measuring the input fields phase difference ($\phi$) at the Lorentzians;
	\item the intrinsic and extrinsic damping rates $\gamma +\Gamma$, by measuring the Full Width at Half Maximum (FWHM) of the Lorentzians.
\end{enumerate}
These parameters are not easily estimated from the usual (one side input) spectral response measurements because, 
the inter-mode interaction strongly modify the energy stored inside the cavity and, furthermore, the minima of the transmission doublet do not directly relate with the real parts of the eigenvalues.

In this subsection, we treated the case of a symmetric interferometric excitation. An asymmetric interferometric excitation induces an asymmetric system response, giving rise to a controlled unbalance of the resonant doublet. It is worth noting that at the diabolic point, i.e. $|\beta|=0$, the response of the bus waveguide/microresonator system is insensitive to any phase variation in the interferometric excitation. In fact, the counter-propagating modes do not interfere, and the output fields show a Lorentzian lineshape.

\begin{figure*}[t!]
	\centering
	\includegraphics[width=1\textwidth]{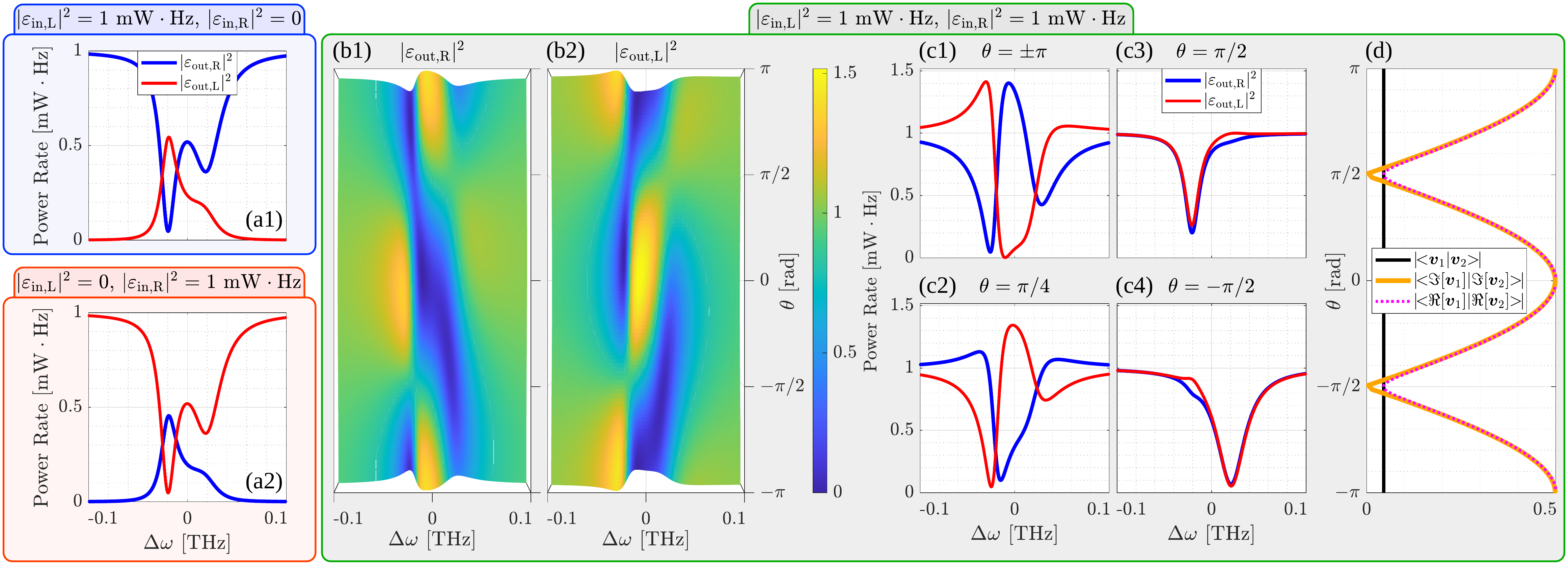}
	\caption{Theoretical results for a non-Hermitian coupling. The transmitted and reflected intensities are shown in panels (a1) and (a2), for left ($\varepsilon_{in,L}=1\,\sqrt{\mWHz}$ and $\varepsilon_{in,R}=0$) or right ($\varepsilon_{in,L}=0$ and $\varepsilon_{in,R}=1\,\sqrt{\mWHz}$) input. The blue (red) curves denote the right (left) output fields. Panels (b1) and (b2) show the intensity of the output fields as a function of the angle $\theta$ and frequency detuning $\Delta \omega$ for a symmetric interferometric excitation ($\varepsilon_{in,L}=\varepsilon_{in,R}=1\,\sqrt{\mWHz}$). (c1)-(c4) Spectral output field intensities for few $\theta$ values given in each panel. (d) Modulus of the eigenstates inner product. The non-Hermitian nature of the coupling makes the eigenstates non-orthogonal, i.e. $\braket{\bm{v}_1|\bm{v}_2} \neq 0$. Here, we used the following coefficients: $\Gamma=\gamma=6.8\GHz$, $\beta_{12}=20.2\GHz$ and $\beta_{21}=(-20.2+9i)\GHz$. 
	%and $\omega_{0}=2\pi \cdot 193\THz$.
	}
	\label{fig:Plot_tot_nonHermitian_th}
\end{figure*}

\subsection{Non-Hermitian coupling and exceptional point}

In this case, there is no relation between $\beta_{12}$ and $\beta_{21}$. Both $h$ and $n$ can be different from zero. Since we are dealing with passive microring, we restrict our-selves to the no-gain case where $|n| \le \gamma$ \cite{Back_Biasi}. As a result, $n$ induces a dissipative term which causes an asymmetric energy exchange between the counter-propagating modes. In the case of a single-side input, an unbalanced (i.e. different intensities) transmission doublet is observed. This is shown in \figsname\ref{fig:Plot_tot_nonHermitian_th} (a1)-(a2). Again, the Lorentz reciprocity theorem ensures equal transmissions \cite{Reciprocity_Potton, Isol_Renner}. On the other hand, the non-Hermitian intermodal coupling induces different reflections for the different input directions \cite{Non-Hermitian_Ashida}.
%%%%%%%
\begin{figure}[b!]
	\centering
	\includegraphics[width=\linewidth]{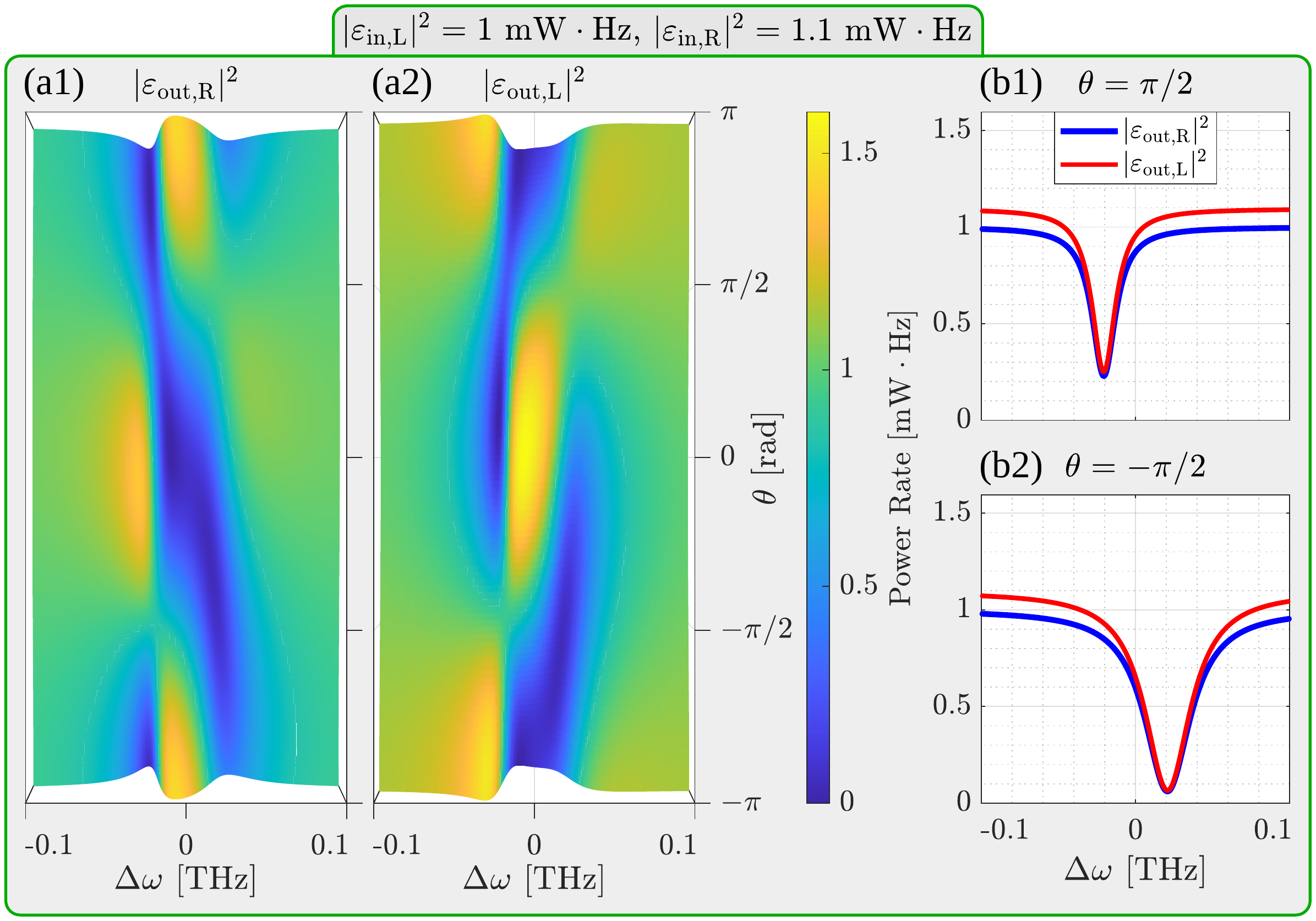}
	\caption{Theoretical results for an asymmetric interferometric excitation, which satisfies \eqsname\eqref{eq:cond_nonHermitian}. Panels (b1) and (b2) show the intensities of the output fields as a function of the angle $\theta$ and frequency detuning $\Delta \omega$. The graphs (b1) and (b2) show the spectral output field intensities for $\theta=\pi/2$ and $\theta=-\pi/2$, respectively. Here, we used the same coefficients of \figname\ref{fig:Plot_tot_nonHermitian_th}, i.e $\Gamma=\gamma=6.8\GHz$, $\beta_{12}=20.2\GHz$ and $\beta_{21}=(-20.2+9i)\GHz$. %$\omega_{0}=2\pi \cdot 193\THz$.
	}
	\label{fig:Plot_tot_nonHermitian_Bilanciato_th}
\end{figure}
%%%%%%

In the case of an interferometric excitation, the non-Hermitian intermodal coupling makes the eigenstates of \eqname\eqref{eq:Eigenvectors_g} non-orthogonal ($\braket{\bm{v}_1|\bm{v}_2} \neq 0$). By writing $\beta_{12/21}=|\beta_{12/21}|\,e^{i\varphi_{12/21}}$, we can write the eigenvalues as:
\begin{equation}
\begin{split}
\lambda_{1,2} =& \omega_{0}\mp\sqrt{|\beta_{12}||\beta_{21}|}\sin{\left[\frac{\varphi_{12}+\varphi_{21}}{2}\right]}+\\
&- i\left(\Gamma+\gamma\mp\sqrt{|\beta_{12}||\beta_{21}|}\cos{\left[\frac{\varphi_{12}+\varphi_{21}}{2}\right]}\right).
\end{split}
\label{eq:EigenvalueSinecosine}
\end{equation}
Here, the real part is related to the splitting of the transmission response. Differently, the imaginary part induces an unbalance in the energy exchange between the counter-propagating modes. In analogy to the Hermitian case, we define the angle $\theta = \phi -\frac{\varphi_{21}-\varphi_{12}}{2}-\frac{\pi}{2}$. It allows establishing a uniform notation and, especially, treating several cases of non-Hermitian coupling.

For a symmetric interferometric excitation, we obtain the maps shown in \figname\ref{fig:Plot_tot_nonHermitian_th} (b1)-(b2). Changing $\phi$, i.e. $\theta$, a re-distribution of energy between the counter-propagating modes is observed, which results in different doublet shapes of the transmitted output fields. For $\theta=\pm \pi$ the dissipative term induces an unbalanced strongly asymmetric doublet (see \figname\ref{fig:Plot_tot_nonHermitian_th} (c1)). For $\theta = \pm \pi/2$, i.e. \figsname\ref{fig:Plot_tot_nonHermitian_th} (c3)-(c4), the doublet merges in a single asymmetric peak, which has not a simple Lorentzian shape as in the Hermitian case. Moreover, $|\braket{\bm{v}_1|\bm{v}_2}|$ is different from zero for each $\theta$, while its real and imaginary parts follow a trend similar to the Hermitian case (\figname\ref{fig:Plot_tot_nonHermitian_th} (d)). However, $|\braket{\Real[\bm{v}_1]| \Real[\bm{v}_2]}|$ does not decrease to zero at $\pm \pi/2$, but shows a minimum equal to $|\braket{\bm{v}_1|\bm{v}_2}|$. In contrast, $|\braket{\Imag[\bm{v}_1]|\Imag[\bm{v}_2]}|$ goes to zero as in the Hermitian case. 
\begin{figure*}[t!]
	\centering
	\includegraphics[width=1\textwidth]{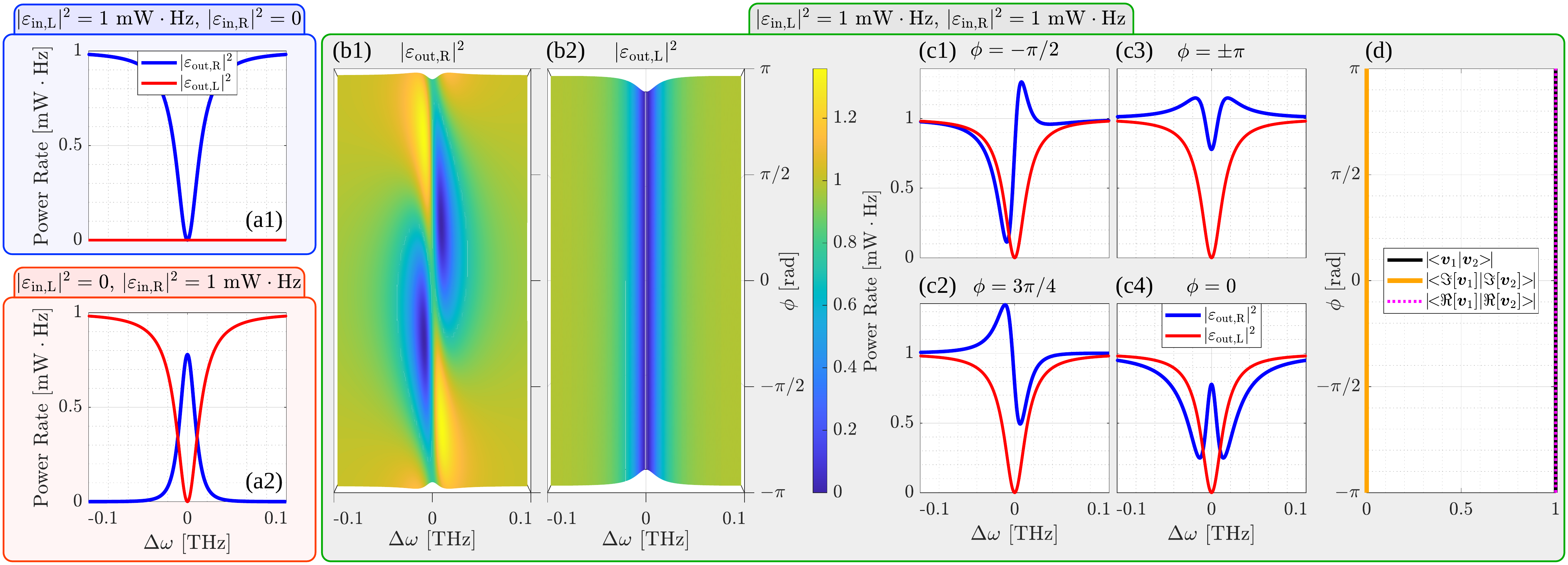}
	\caption{Theoretical results for a taiji microresonator coupled to a bus waveguide. Panels (a1) and (a2) show the transmitted and reflected intensities as a function of the frequency detuning $\Delta\omega$ at the left and right outputs, respectively. The blue (red) curves refer to the output fields from the right (left) waveguide edges. The maps (b1) and (b2) show the intensities of the output fields as a function the phase $\phi$ and $\Delta\omega$ for a symmetric interferometric excitation. The transmitted spectral intensities are plotted on the graphs (c1)-(c4) for $\phi=-\pi/2$, $3\pi/4$, $\pm\pi$ and $0$, respectively. The graph (d) shows the modulus of the inner product of the eigenstates. As expected, the taiji works on an exceptional point, and therefore, the eigenstates are parallel, i.e $\braket{\bm{v}_1|\bm{v}_2}=1$. Here, we used the following coefficients: $\Gamma=\gamma=6.8\GHz$, $\beta_{12}=12\GHz<2\gamma$ and $\beta_{21}=0$. 
	%and $\omega_{0}=2\pi \cdot 193\THz$.
	}
	\label{fig:Plot_tot_nonHermitian_Taiji_th}
\end{figure*}
%%%%%%%

The lack of coalescence of the doublet into a simple Lorentzian, when both the real and imaginary parts of the eigenstates reach their inner product minima, is related to the dissipative part of the coupling ($n$). To get a Lorentzian, we can use an asymmetric interferometric excitation. Indeed, when
\begin{equation}\label{eq:cond_nonHermitian}
\varepsilon_{\rm in,R}=\sqrt{\frac{|\beta_{21}|}{|\beta_{12}|}}\varepsilon_{\rm in,L}\quad\&\quad \theta=\pm\frac{\pi}{2}+2\pi m\,,
\end{equation}
where $m\in\mathbb{Z}$, the transmission lineshapes reduce to a single Lorentzian lineshape (\figname\ref{fig:Plot_tot_nonHermitian_Bilanciato_th}). The first condition in \eqname\eqref{eq:cond_nonHermitian} compensates for the different strengths of the coupling coefficients, while the second ensures the same interference between the transmitted and reflected fields for both directions. As a result, \eqname\eqref{eq:cond_nonHermitian} establishes the proportionality between the output field intensities. Substituting \eqname\eqref{eq:cond_nonHermitian} in the \eqname\eqref{eq:tR}, we obtain:
\begin{gather}
\label{eq:LorentzianNonHermitian}
\begin{pmatrix}
|\varepsilon_{\rm out,R}|^{2}  \\
|\varepsilon_{\rm out,L}|^{2}
\end{pmatrix}\!\!
=\!\! \left(\!\! 1-\frac{4(\gamma\mp\tilde{\gamma})\Gamma}{(\Delta\omega\mp\tilde{\beta})^2+(\gamma\mp\tilde{\gamma}+\Gamma)^2}\!\!\right)\!\!
\begin{pmatrix}
|\varepsilon_{\rm in,L}|^{2}  \\
|\varepsilon_{\rm in,R}|^{2}
\end{pmatrix},\\
\tilde{\beta}:=\sqrt{|\beta_{12}||\beta_{21}|}\,\sin{[(\varphi_{12}+\varphi_{21})/2]}\,,\nonumber\\
\tilde{\gamma}:=\sqrt{|\beta_{12}||\beta_{21}|}\,\cos{[(\varphi_{12}+\varphi_{21})/2]}\,.\nonumber
\end{gather}
$\tilde{\beta}$ and $\tilde{\gamma}$ are the real and imaginary parts of the shifts of the eigenvalues with respect to $\omega_{0}$ (see \eqname\eqref{eq:EigenvalueSinecosine}).

\figsname\ref{fig:Plot_tot_nonHermitian_Bilanciato_th} (a1)-(a2) show the maps of the transmitted field intensities as a function of $\Delta \omega$ and $\theta$ for an asymmetric interferometric excitation which satisfies (\ref{eq:cond_nonHermitian}). For $\theta = \pm \pi/2$  (\figname\ref{fig:Plot_tot_nonHermitian_Bilanciato_th} (b1)-(b2)), the transmitted fields show a simple Lorentzian lineshape, and most importantly, their minima indicate the real part of the eigenvalues. Furthermore, the FWHM of the Lorentzian lineshapes are related to the imaginary part of the eigenvalues and depend on the intrinsic microresonator losses. Thus, the interferometric excitation allows studying both the real and imaginary parts of the eigenvalues. In (\figname\ref{fig:Plot_tot_nonHermitian_Bilanciato_th} (b1)) the best quality factor of the microresonator in the absence of backscattering corresponds to the value obtained by setting the relative phase $\phi$ so that $\theta = \pi/2$.

%%%%%%% posizione originale
%\begin{figure}[t!]
%    \centering
%    \includegraphics[width=\linewidth]{Figures/Plot_tot_nonHermitian_Bilanciato_th.pdf}
%    \caption{Theoretical results for an asymmetric interferometric excitation, which satisfies \eqsname\ref{eq:cond_nonHermitian}. Panels (b1) and (b2) show the intensities of the output fields as a function of the phase and frequency detuning. The graphs (b1) and (b2) display their cross-section for a fixed phase equal to $\pi/2$ and $-\pi/2$, respectively. The two output fields coalesce to a single Lorentzian exhibiting the characteristic of the eigenvalues. Here, we used the same coefficients of \figname\ref{fig:Plot_tot_nonHermitian_th}, i.e $\Gamma=\gamma=6.8\GHz$, $\beta_{12}=20.2\GHz$ and $\beta_{21}=(-20.2+9i)\GHz$.}
%    \label{fig:Plot_tot_nonHermitian_Bilanciato_th}
%\end{figure}
%%%%%%

When just one of the two coupling coefficients $\beta_{12/21}$ tends to zero, the eigenvalues (\eqname\eqref{eq:EigenvalueSinecosine}) vary following a square root dependence. More importantly, when one of the two coefficients is zero, the eigenvalues become degenerate and the eigenstates (\eqname\eqref{eq:Eigenvectors_g}) become parallel ($\braket{\bm{v}_1|\bm{v}_2}=1$). Therefore, the system works on an exceptional point \cite{Degeneracy_Berry}. This situation might be found in an ideal taiji microresonator (\figname\ref{fig:microresonators} (b)). Neglecting the backscattering, this microresonator has $\beta_{12}\neq 0$ and $\beta_{21}=0$. In this case, the energy exchange between the counter-propagating modes is strongly unbalanced in one direction. As can be seen in panels (a1)-(a2) of \figname\ref{fig:Plot_tot_nonHermitian_Taiji_th}, the left- or right-side transmission shows the simple Lorentzian lineshape when a single input excitation is used. On the other hand, the two reflections are completely different due to the non-Hermitian intermodal coupling. The intensity reflected from the right waveguide edge shows a peak, while that from the left waveguide edge is zero. The response of the bus waveguide/taiji microresonator system for a symmetric interferometric excitation is shown in panels (b1)-(b2) of \figname\ref{fig:Plot_tot_nonHermitian_Taiji_th}.  It is observed that, unlike the diabolic point case, the exceptional point case is affected by an interferometric excitation. In particular, the left output field is unaffected by a $\phi$ change and maintains the simple Lorentzian lineshape. While, the right output field has a lineshape that depends markedly on $\phi$ (\figname\ref{fig:Plot_tot_nonHermitian_Taiji_th} (c1)-(c4)). For $\phi=0$, the right output field intensity exhibits a balanced doublet (\figname\ref{fig:Plot_tot_nonHermitian_Taiji_th} (c4)). \figname\ref{fig:Plot_tot_nonHermitian_Taiji_th} (d) shows the inner products of the eigenstates. As expected for a microresonator at the exceptional point, they do not depend on $\phi$. In particular, the inner products of the eigenstates and of their real parts are equal to $1$, while they are zero for the imaginary ones.

\begin{figure*}[t!]
	\centering
	\includegraphics[width=1\textwidth]{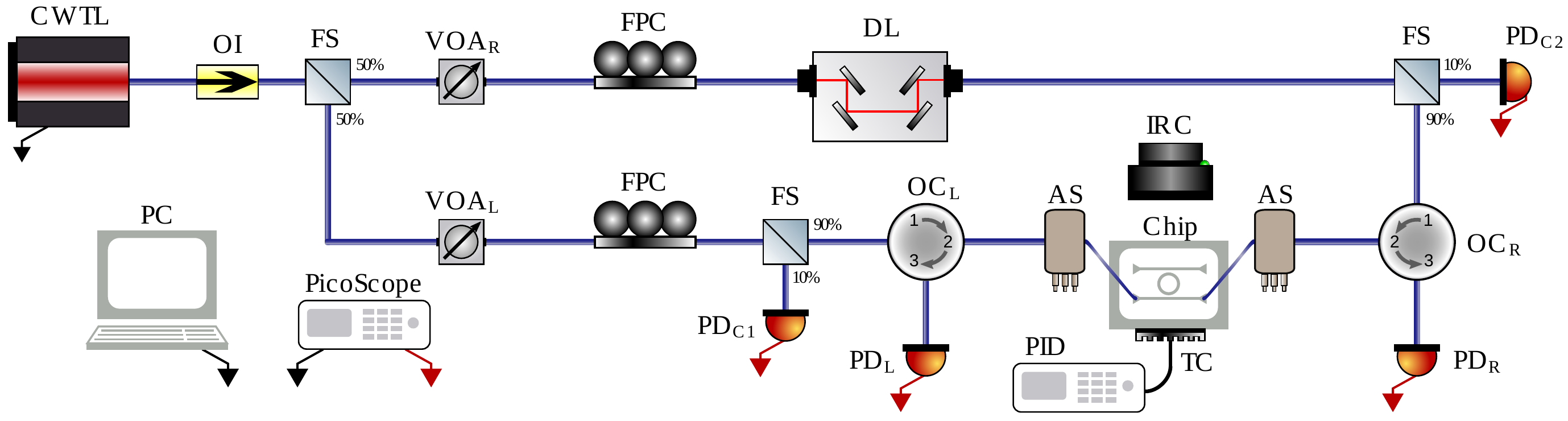}
	\caption{Sketch of the experimental setup. PC: \textit{Personal Computer} to control the instruments, CWTL: \textit{Continuous Wave Tunable Laser} to excite the system, OI: \textit{Optical Isolator} to protect the Laser from backreflations, FS: \textit{Fiber Splitter} to split the signal, VOA: \textit{Variable Optical Attenuator} to control the amplitudes, FPC: \textit{Fiber Polarization Controller} to set the Transverse Electric polarization at the chip grating couplers, DL: \textit{Delay Line} to balance the two optical paths, OC: \textit{Optical Circulator} to measure and excite simultaneously the counter-propagating fields, PD: \textit{Photo Detector} and PicoScope \textit{PC Oscilloscope} to measure the output powers, AS: \textit{Alignment Stage} and IRC: \textit{Infrared Camera} to see and align the stripped fibers with the sample structures, TC: \textit{Temperature Controller} and PID \textit{Proportional–Integral–Derivative controller} to fix the chip temperature.}
	\label{fig:Setup}
\end{figure*}

As a last comment, \eqname\eqref{eq:EigenvalueSinecosine} establishes that it is possible to reach the exceptional point in several ways. In fact, by vanishing to zero one of the two intermodal coupling coefficients, e.g. $\beta_{21} \rightarrow 0$ while $\beta_{12}\neq0$, one can have different situations depending on the linear combination of their relative phases ($\frac{\varphi_{12}+\varphi_{21}}{2}$). Precisely, when $\frac{\varphi_{12}+\varphi_{21}}{2}=\pi/2+m \,\pi$ with $m \in \mathbb{Z}$, the degeneracy is achieved as a result of the change of just the real (resonant) part. Conversely, when $\frac{\varphi_{12}+\varphi_{21}}{2}= m \,\pi$ with $m \in \mathbb{Z}$, the two eigenvalues coalesce as a consequence of the change of just the imaginary (absorptive) part. Whereas for any other phases, an intermediate situation is obtained, and the degeneracy is achieved in the complex plane of the eigenvalues. Thus, the phase relation between  $\beta_{12}$ and $\beta_{21}$ allows to study an exceptional point which can be resonant, absorptive or both, in the same way as in the recent work \cite{Exceptional_Lan}. From \eqname\eqref{eq:EigenvalueSinecosine}, it follows that it is convenient to use the phase relation that makes the perturbation on the eigenvalues purely real in the case of sensing through passive microresonators capable of working on an exceptional point. This permits to estimate the strength of the perturbation by simply measuring the splitting in the transmission spectra \cite{Sensing_Wiersig,Sensing_Lan}.

%%%%%%% posizione originale
%\begin{figure*}[t!]
%    \centering
%    %\includegraphics[width=1\textwidth]{Figures/Plot_tot_nonHermitian_Taiji_th_old01.pdf}
%    \includegraphics[width=1\textwidth]{Figures/Plot_tot_nonHermitian_Taiji_th.pdf}
%    \caption{Theoretical results for a taiji microresonator coupled to a bus waveguide. Here, $\beta_{12} \neq 0$ and $\beta_{21}=0$. Panels (a1) and (a2) show the transmitted and reflected intensities as a function of the frequency detuning exiting the system from left and right, respectively. The blue (red) curves highlights the output field from the right (left) facet. The maps (b1) and (b2) show the intensities of the output fields to a symmetric interferometric excitation as a function of both phase and frequency detuning. Their cross-sections are plotted on the graphs (c1)-(c4) for $-\pi/2$, $3\pi/4$, $\pm\pi$ and $0$, respectively. The graph (d) shows the modulus of the inner product of the eigenstates. As expected, the taiji works on an exceptional point, and therefore, the eigenstates are parallel, i.e $\braket{\bm{v}_1|\bm{v}_2}=1$. Here we used the following coefficients: $\Gamma=\gamma=6.8\GHz$, $\beta_{12}=12\GHz<2\gamma$, $\beta_{21}=0\GHz$.}
%    \label{fig:Plot_tot_nonHermitian_Taiji_th}
%\end{figure*}
%%%%%%%

\section{Experimental measurements}
\label{sec:Exp}

\subsection{Experimental setup and samples}

Our setup is sketched in \figname\ref{fig:Setup}. A fiber-coupled continuous wave tunable laser (Yenista OPTICS, TUNICS-T100S) is used as the source. It operates at $2\, \rm{mW}$ spanning a range of wavelength from $1470\,\rm{nm}$ to $1580\,\rm{nm}$. Its emission passes through an optical isolator to avoid any spurious back-reflection, and then, through a $50:50$ fiber splitter. This allows the coherent interferometric excitation. After the fiber splitter, the two coherent signals in the different fibers are modulated in intensities by a variable optical attenuator and are adjusted in polarization by a polarization control stage. We use the transverse electric (TE) polarization in the waveguide. In order to balance the two optical paths, the signal of one fiber is sent to a free space delay line. Two fiber taps of $10\%$ connected to two photodiodes (Thorlabs, PDA10CS2) are used to monitor the input signals. After these, two optical circulators route these signals into two single mode stripped fibers that couple the light into the sample. As clear in the figure, the sample is excited from both the left and right sides by two coherent light signals. The output fields are collected by the same stripped fibers and are routed by the circulator into  two other photodiode detectors (Thorlabs, PDA10CS2), labeled $\rm{PD_L}$ and $\rm{PD_R}$ in \figname\ref{fig:Setup}. Finally, an oscilloscope (PicoScope 4000 Series) records simultaneously the signals of all the four detectors. The sample is mounted on a thermostated holder whose temperature is controlled by a Proportional-Integral-Derivative controller (SIM960 Analog PID controller) connected to a peltier cell and a $10\kOhm$ thermistor. The same setup is also used for standard (single side excitation) measurements. In these, the signal is input from the left or right only.

Two different microresonator geometries have been studied, which are based on silicon waveguides with a cross section of $450\nm \times 220\nm$ embedded in silica cladding. The first microresonator geometry has a ring shape with a radius of $7\um$ and it is point-like coupled to two bus waveguides in the common add and drop configuration with a gap of $300\nm$. These  devices are described in detail in \cite{Rings_Borghi}. The second is a taiji microresonator with a racetrack geometry. It is obtained by coupling four Euler curves (radius $15\um$ and angle $90^\circ$) to two pairs of straight sections of length $758\nm$ and $3106\nm$. The perimeter of the microresonator is about $196.22 \um$. The inner S-shaped waveguide is designed by using four Euler curves. The central ones have a radius of $10 \um$ and an angle of $135^\circ$ and those at the ends have a radius of $10 \um$ and an angle of $75^\circ$. At both ends of the S-shaped waveguide, an inverse tapering ensures no back-reflections. The gaps between the taiji microresonator and the S-shaped waveguide are $210\nm$. The one between the taiji and the bus waveguide is $241\nm$. In both types of microresonators, grating couplers ($\simeq \,3.7\dB$ coupling loss) are used to input and output the light signals. The microresonators have been fabricated in different runs at the IMEC/Europratice facility within the multi-project wafer program.

\subsection{Method to control the phase shift}

Since the interferometric excitation relies on coupling the light signal in both propagation directions, spurious reflections can induce unwanted interference fringes in the spectral measurements. In order to reduce these, the weak Fabry-Perot oscillations due to the reflections by the grating couplers \cite{Gratings_Marchetti} has been decreased by using glycerol as a matched index fluid. Furthermore, the different optical paths of the fibers after the first 50:50 beam splitter are compensated by a delay line. In this way, the free spectral range (i.e. period) of the interference fringes caused by the large interferometric setup is much wider than a single microresonator resonance doublet. 

In our setup, we do not have an absolute control of the interferometric excitation phase ($\phi$). We leave $\phi$ to vary randomly and we extract its value from a fit of the spectra. In fact, the phase shift between the two splitted signals is affected by fiber relaxation, air fluxes and room temperature variations whose accurate control is difficult. However, the nearly-balanced condition makes the time of this random phase variations much longer than the time of a single spectral acquisition. Specifically, the phase variation rate is measured to be $\frac{\diff\phi}{\diff t}\le 0.1~\rm rad/s$, while a typical acquisition time for a $\Delta \lambda =0.3\nm$ wide spectrum is shorter than $\Delta\lambda/\frac{\diff\lambda}{\diff t} < (0.3\nm) /(1~\rm{nm/s}) = 0.3~\rm s$, where $\frac{\diff\lambda}{\diff t}$ is the laser wavelength scanning rate. Hence, the phase variation during a single scan is less than $3\times10^{-2}~{\rm rad}$, and, therefore, we can safely assume it negligible. A scanning rate of $1~\rm{nm/s}$ is permitted by the use of the Picoscope, which synchronizes the laser wavelength scan with the recording of the data from the photodetectors. Furthermore, the symmetrical excitation is made possible by the calibration of all detectors, the measurement of the propagation losses along the two paths, and the estimation of the coupling losses of the gratings (see also \appname\ref{app:exp}). The measurement of the resonance lineshapes, in the case of interferometric excitation, has been obtained by acquiring 100 output fields spectra. Each spectrum has a resolution of $0.1~{\rm pm}$. Furthermore, a guard time of 1 second has been waited between each acquisition to ensure a sufficiently large relative phase variation.

\subsection{Results}
\subsubsection{Microring resonator}

\begin{figure*}[t!]
	\centering
	\includegraphics[width=1\textwidth]{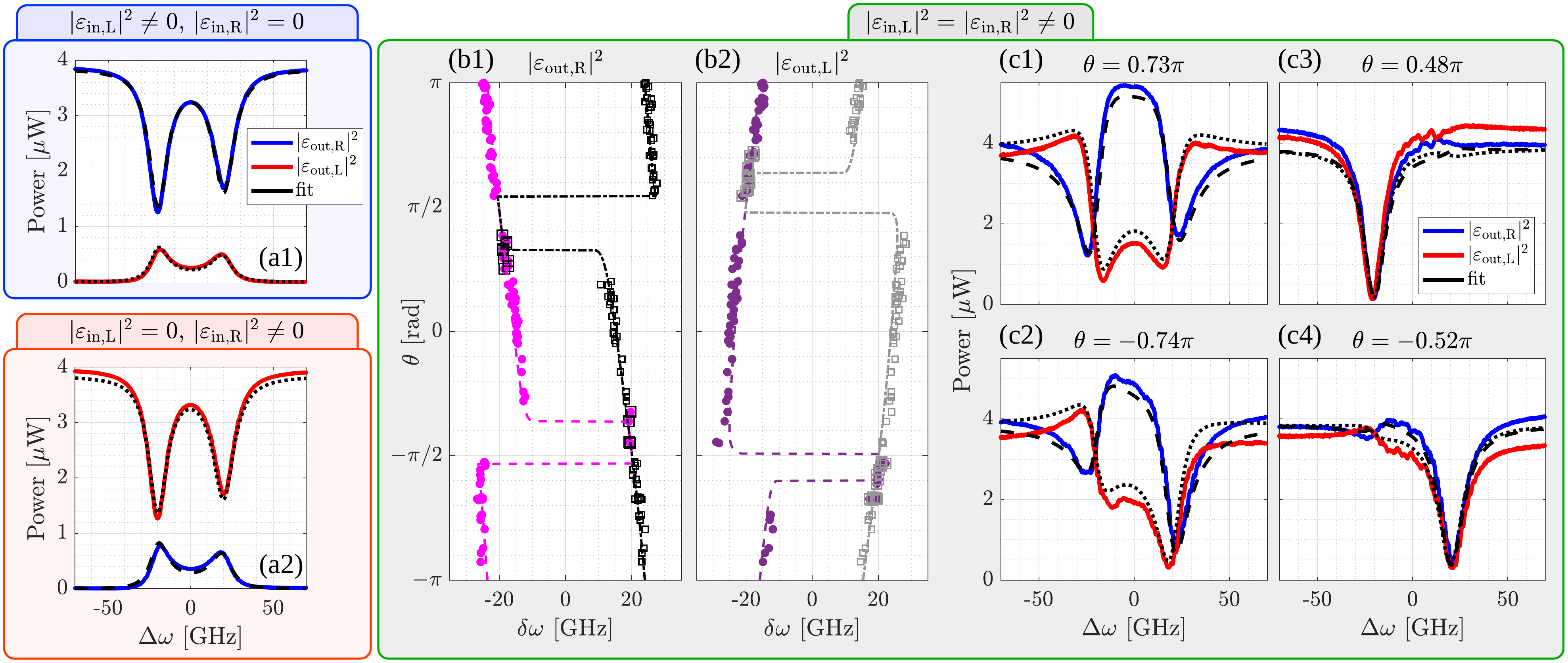}
	\caption{Experimental results for a microring. Panels (a1) and (a2) show the transmitted and reflected intensities as a function of the frequency detuning ($\Delta \omega$) for the right and left excitation. Panels (b1) and (b2) display the doublet splitting ($\delta \omega=\omega_{</>}-\omega_{0}$) for a symmetric interferometric excitation as a function of the phase $\theta$ for the right or the left output fields. $\omega_{</>}$ are the measured frequency minima of the doublet. The magenta (purple) dots refer to $\delta \omega=\omega_{<}-\omega_{0}$ for the right (left) output fields. The black (gray) squares refer to $\delta \omega=\omega_{>}-\omega_{0}$ for the right (left) output fields. The dashed magenta (purple) and dash-dotted black (gray) curves show the theoretical results modeled by using the fitting parameters of the one-side excitation. Panels (c1), (c2), (c3) and (c4) show the output field intensities as a function of $\Delta \omega$ for $\theta=0.73\,\pi$, $-0.74\,\pi$, $0.48\,\pi$ and $-0.52\,\pi$, respectively. The blue (red) curves refer to the right (left) output field intensities. The dashed-black and dotted-black lines are the fit with the theoretical model for the right and left output fields, respectively. From the fit we obtain the following coefficients: $\Gamma= (2.662 \pm 0.003)\GHz$, $\gamma = (4.56 \pm 0.01)\GHz$, $\beta_{12} = [(-19.72 \pm 0.02) - i (0.2 \pm 0.4)]\GHz$, $\beta_{21} = [(20.67 \pm 0.02) + i (0.8 \pm 0.4)]\GHz$.
	}
	\label{fig:Plot_tot_nonHermitian_exp}
\end{figure*}

The experimental results for the microring resonator are reported in \figname\ref{fig:Plot_tot_nonHermitian_exp}. Panels (a1) and (a2) show the transmitted and reflected intensities for a single left and right excitation, respectively. It can be observed that the backscattering due to the surface-wall roughness gives rise to a slightly unbalanced doublet. The left (blue line in the top panel) and right (red line in the bottom panel) transmission spectra are equal, as a consequence of the Lorentz reciprocity theorem. On the contrary, the left (red line) and right (blue line) reflections show similar features but different intensities. This is due to the non-Hermitian intermodal coupling, which induces a different energy exchange between the counter-propagating modes. The fit with the theoretical model (\eqname\eqref{eq:tR} and \eqname\eqref{eq:tL} in the single side excitation) is shown by the dashed-black lines (right output fields) and the dotted-black lines (left output fields).
The fit yields the resonance frequency ($\omega_{0}\simeq 2\pi \cdot 196.593\THz$) and the intrinsic ($\gamma = (4.56 \pm 0.01)\GHz$), the extrinsic ($\Gamma= (2.662 \pm 0.003)\GHz$), and the intermodal coupling ($\beta_{12} = [(-19.72 \pm 0.02) - i (0.2 \pm 0.4)]\GHz$ and $\beta_{21} = [(20.67 \pm 0.02) + i (0.8 \pm 0.4)]\GHz$) coefficients.
As a result, the non-Hermitian coefficient $n = [(0.48\pm0.01) - i (0.5\pm0.3)]\GHz$ is smaller than the Hermitian coefficient $h = [(-0.3\pm0.3) -i (20.20\pm0.01)]\GHz$. The effect of $n \neq 0$ is visible in the unbalanced doublet because of the relative high Q-factor.

\begin{figure*}[t!]
	\centering
	\includegraphics[width=0.8221\textwidth]{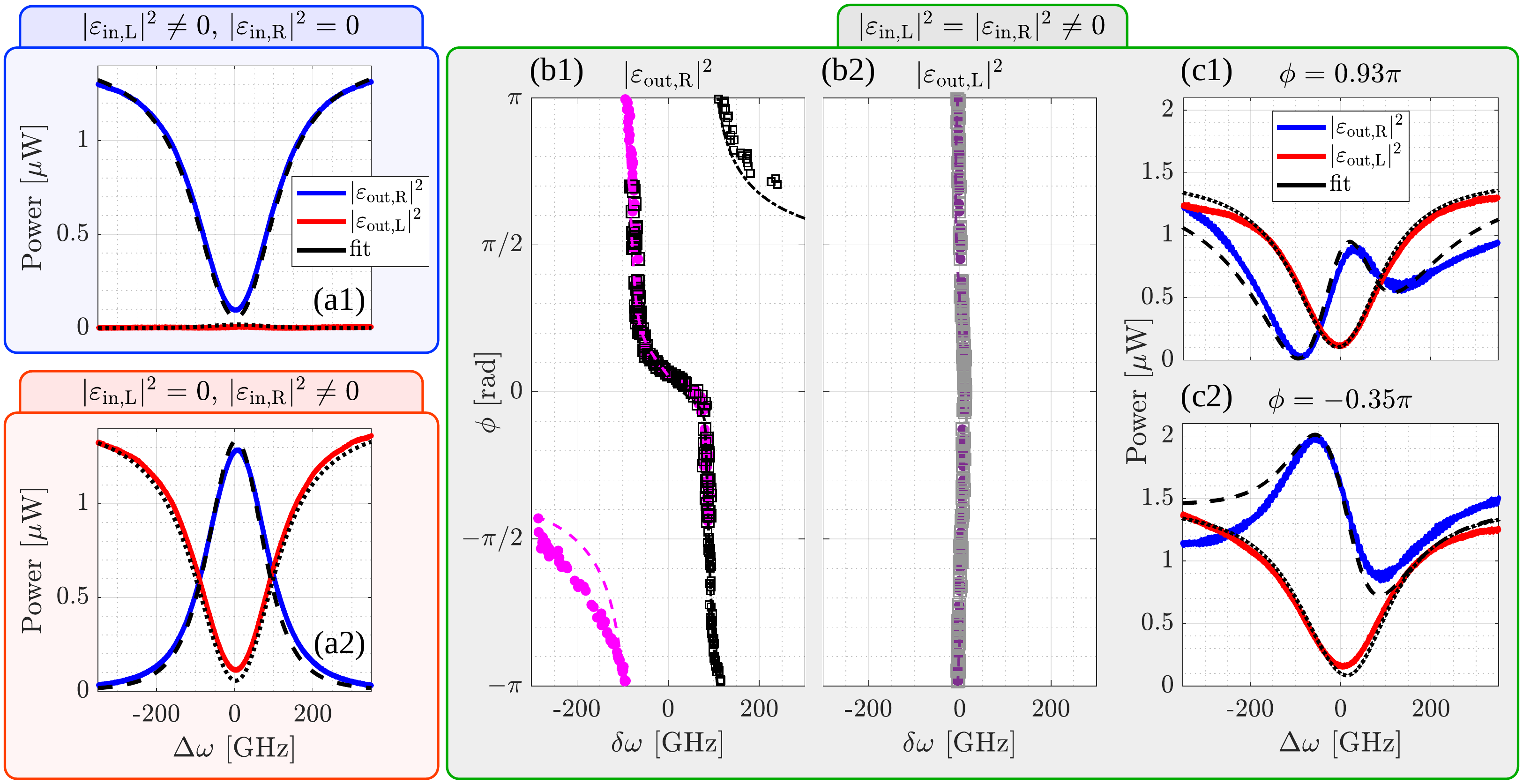}
	\caption{Experimental results for a taiji microresonator. The red (blue) curves refer to the left (right) output field. The dashed and dash-dotted black lines show the fit with the theoretical model. Panels (a1) and (a2) show the output fields as a function of the frequency detuning $\Delta \omega$ for a left and right excitation, respectively. Panels (b1) and (b2) display the doublet splitting ($\delta \omega=\omega_{</>}-\omega_{0}$) for a symmetric interferometric excitation as a function of the phase $\phi$ for the right or the left output fields. $\omega_{</>}$ are the measured frequency minima of the doublet. The magenta (purple) dots refer to $\delta \omega=\omega_{<}-\omega_{0}$ for the right (left) output fields. The black (gray) squares refer to $\delta \omega=\omega_{>}-\omega_{0}$ for the right (left) output fields. The dashed magenta (purple) and dash-dotted black (gray) curves show the theoretical results modeled by using the fitting parameters of the one-side excitation. Panels (c1) and (c2) show two example of spectra for an interferometric excitation with a phase $\phi = 0.93\pi$ and $-0.35\pi$, respectively. From the fit we obtain the following coefficients: $\Gamma= (75.1 \pm 0.1)\GHz$, $\gamma = (42.6\pm 0.1)\GHz$, $\beta_{12} = [(-90.8\pm 0.1) -i(24.17\pm 0.03)]\GHz$, $\beta_{21} = [(8.7\pm 0.1) -i(7.30\pm 0.09)]\GHz$.
	}
	\label{fig:Plot_tot_taiji_exp}
\end{figure*}

Measurement results by the symmetric interferometric excitation are summarized in \figname\ref{fig:Plot_tot_nonHermitian_exp}. Panels (b1) and (b2) show the doublet splitting as a function of $\phi$. Since our setup does not allow an absolute control of $\phi$, $\theta = \phi -\frac{\varphi_{21}-\varphi_{12}}{2}-\frac{\pi}{2}$ was estimated by fitting the 100 experimental spectra of the interferometric excitation with the model of \eqname\eqref{eq:tR} and \eqname\eqref{eq:tL}. In this, we used the coefficients estimated for the single side excitation as fixed parameters and the relative phase $\phi$ as the only free fit parameter. Typically the error on $\phi$ is of $3\times10^{-2}~{\rm rad}$.
In the figure, the doublet splitting is reported as a frequency detuning with respect to the microring resonance $\delta \omega = \omega_{</>} - \omega_0$, where $\omega_{</>}$ are the measured frequencies of the doublet minima and $</>$ refer to the position with respect to $\omega_{0}$. Panels (b1) and (b2) report $\delta \omega$ for the right and left output fields, respectively. In the plot, we use the following notation: the magenta (purple) dots refer to $\delta \omega= \omega_< - \omega_0$, the black (gray) squares refer to $\delta \omega= \omega_> - \omega_0$, the dashed magenta (purple) and dash-dotted black (gray) curves show the theoretical calculated detuning by \eqsname\eqref{eq:tR} and (\ref{eq:tL}) with the parameters from the single-side excitation fits. It worth noticing that the analytical model of Sec. \ref{sec:Theory} reproduces the experimental data. 

\figname\ref{fig:Plot_tot_nonHermitian_exp} (c1)-(c4) show the intensities of the output fields as a function of $\Delta \omega$ for $\theta = 0.73\,\pi$, $-0.74\,\pi$, $0.48\,\pi$, and $-0.52\,\pi$. Notation is the same as for panels (a1) and (a2). It is clear that our theoretical model is in agreement with the experimental data. As expected, the $\phi$ variation induces a deformation of the doublet which, when $\theta \simeq \pm \, \pi/2$, merges into a single resonance. For these $\theta$ values, the dots coincide with the square in panels (b1) and (b2). Specifically, graphs (c3) and (c4) show that the output fields have a quasi-simple Lorentzian lineshape. Their resonant frequencies correspond to the real part of the eigenvalues with a negative (panel (c3)) and positive (panel (c4)) detuning.

\begin{table}[b!]
	\begin{tabularx}{0.47\textwidth} { 
			| >{\raggedright\arraybackslash}X 
			| >{\centering\arraybackslash}X 
			| >{\centering\arraybackslash}X | }
		\hline
		& Single Side Excitation $[\GHz]$ & Interferometric Exitation $[\GHz]$ \\ \hline
		$|\Real[\lambda_{1,2}-\omega_0]|$ & $20.2 \pm 0.3$ & $21.0 \pm 0.3$ \\ \hline
		$\Imag[\lambda_{1}]$ & $-6.73 \pm 0.02$ & $-6.8 \pm 0.2$ \\ \hline
		$\Imag[\lambda_{2}]$ & $-7.71 \pm 0.02$ & $-7.51 \pm 0.07$ \\ \hline
	\end{tabularx}
	\caption{Real and imaginary part of the microresonator’s eigenvalues. The single side excitation column shows the values calculated by using the parameters obtained from the fit of the spectral transmission and reflection responses (panels (a1) and (a2) of \figname\ref{fig:Plot_tot_nonHermitian_exp}). The column labelled interferometric excitation shows the values estimated by using the experimental data in panels (c3)-(c4) of \figname\ref{fig:Plot_tot_nonHermitian_exp}. }
	\label{tabel}
\end{table}

Assuming a pure Hermitian case, we can estimate from the spectra the coefficient parameters following the method described in Sec. \ref{sec:Theory}.
Table \ref{tabel} shows the comparison between the real and imaginary parts of the eigenvalues $\lambda_{1,2}$ extracted by a full fit of the single side spectra (\figname\ref{fig:Plot_tot_nonHermitian_exp} (a1)-(a2)) or estimated from the spectral positions of the collapsed doublet (\figname\ref{fig:Plot_tot_nonHermitian_exp} (c3)-c(4)). In particular, we fit the interferometric excitation spectra with a Lorentzian lineshape to extract the spectral positions and the FWHM of the merged doublets. Table \ref{tabel} shows that the two methods yield comparable results. Note that the different values and the large error bars for the estimated  $\Imag[\lambda_{1,2}]$ are due to the presence of a non-Hermitian term $n\neq0$ (see \eqname\eqref{eq:EigenvalueSinecosine}). In fact, the slightly non-Hermitianity, the symmetric interferometric excitation and a $\theta \neq \pm \pi/2$, causes a weak perturbation to the Lorentzian lineshape which is riflected in the values given in Table \ref{tabel} (see also \figname\ref{fig:Plot_tot_nonHermitian_th}). From the FWHM of the Lorentzian lineshapes, it is still possible to estimate the Q-factor of the microring in absence of surface-wall roughness which results to be $\simeq 9\times10^{4}$.

\subsubsection{Taiji microresonator}

The experimental measurements for the taiji microresonator are shown in \figname\ref{fig:Plot_tot_taiji_exp}. As in the case of the microring, we first characterized the taiji microresonator with single side excitation experiments. Results are reported in the panels (a1) and (a2). In agreement with the Lorentz reciprocity theorem, the transmitted intensities are equal. In contrast, the presence of the inner S-shaped waveguide makes the reflection approximately equal to zero for left excitation and close to one for right excitation.
Fitting the spectra with the \eqname\eqref{eq:tR} and \eqname\eqref{eq:tL} in the single side excitation yields the values of the resonance frequency ($\omega_{0}\simeq 2\pi \cdot 194.430\THz$) and the intrinsic ($\gamma = (42.6\pm 0.1)\GHz$), the extrinsic ($\Gamma= (75.1 \pm 0.1)\GHz$), and the intermodal coupling ($\beta_{12} = [(-90.8\pm 0.1) -i(24.17\pm 0.03)]\GHz$ and $\beta_{21} = [(8.7\pm 0.1) -i(7.30\pm 0.09)]\GHz$) coefficients.
The coefficient $\beta_{21}$ is non-zero due to the presence of surface-wall roughness which causes a weak backscattering. The reflection due to backscattering is about $8.5 \times 10^{-3}$ times less intense than the reflection due to the inner S-shaped waveguide. Panels (b1) and (b2) show the doublet splitting for a symmetric interferometric excitation as a function of the $\phi$ and $\delta \omega$. As before, $\phi$ is estimated by fitting 100 spectra with only the relative phase as a free parameter while all the other coefficients are equal to the ones estimated from the single side experiments. Noteworthy, the theory shown in \figname\ref{fig:Plot_tot_nonHermitian_Taiji_th} catches the experimental results. The doublet is only observed in the right output field, while the left output field is insensitive to the $\phi$ variations. The panel (b1) shows that the right output field exhibits a clear doublet for a relative phase of about $\pm \pi$. This doublet is observed until $\phi =\pm \pi/2$ where it merges into a single peak.
In this case, the dots and squares perfectly overlap, and in the region between $-\pi/2$ and $\pi/2$, the peak resonant frequency shifts from the right to the left of $\omega_0$. On the contrary, as shown in panel (b2), the left output field does not exhibit any splitting. The dots and squares overlap for all $\phi$ values. As a result, the spectrum always shows a single Lorentzian. However, the presence of the small coefficient $\beta_{21} \neq 0$ gives rise to small oscillations of the resonant frequency. As an example, two experimental spectra and their theoretical fits are shown in panels (c1) and (c2) for $\phi = 0.93 \pi$ and $0.35 \pi$, respectively. Noteworthy, also in this case the model of Sec. \ref{sec:Theory} is in perfect agreement with the experimental results.

\section{Conclusion}
\label{sec:Con}

We theoretically model and experimentally validate the coherent interferometric excitation technique as a method to estimate the real and imaginary parts of the eigenvalues of a Hermitian and non-Hermitian two-level optical system. We applied it to the measurement of the transmission spectra of a bus waveguide/microresonator system where the microresonator is characterized by two degenerate counter-propagating optical modes. The variation of the relative phase and amplitude of the input fields causes the merging of the typical resonant doublet due to the counter-propagating microresonator modes into a single Lorentzian lineshape. This allows a direct estimation of the intermodal coupling coefficients from the transmission spectra in the Hermitian case. In the non-Hermitian case, it allows extrapolating the real part of the eigenvalues from the resonant frequencies and its imaginary part from the resonance shape, i.e., the FWHM.  In particular, it is possible to estimate also the microresonator Q-factor in absence of the intermodal coupling between the counter-propagating modes, i.e. without the contribution of surface-wall roughness backscattering. Furthermore, we show that a taiji microresonator works on an exceptional point and that it has an asymmetric transmission under a coherent interferometric excitation. In fact, the transmitted output field from one side remains unaffected by changes in the relative input field phases, while the other transmitted field exhibits a rich dynamics with a resonance splitting at characteristic $\phi$. These differences are due to the presence of the inner S-shaped waveguide to which the propagating field couples  only in one direction. 

These results represent a proof of concept of the coherent interferometric excitation technique. Indeed, the experimental setup is made of discrete optical components and lacks an absolute control of $\phi$. However, the coherent interferometric excitation can be easily integrated in a single chip, where phase shifters, splitters and integrated Mach-Zehnders can controll both the phase and the amplitudes of the input fields to excite the microresonator counter-propagating modes, see e.g. \cite{Mancinelli2013}. For this reason, our work paves the way to a family of devices based on the exceptional points for sensing applications where the accuracy of the spectral measurement is improved by the coalescence of the eigenvalues.

Finally, the proposed method and analysis can be also generalized to study other two level optical systems such as two coupled micoresonators or photonic molecule \cite{Boriskina_Molecule, Borghi_Molecul} where the complex interplay of different optical modes is difficult to unfold. In particular, the integration of a controlled coherent interferometric excitation into these schemes leads to study different kinds of degeneracies, which can be engineered in the spectrum of coupled optical microcavities \cite{Exceptional_Lan}. As a result, the spectral response of to an interferometric excitation allows to directly identify purely absorptive or radiative exceptional points.

\appendix

\section{Hermitan coupling} 
\label{app:Hermitian}
In this section, we show the Hermitian condition on the coupling coefficients and how it reduces the other characteristic parameters such as $g_{1,2}$, the eigenvalues and the eigenvectors.

From \eqname\eqref{eq:TCMT} of the main text we define the following backscattering matrix:
\begin{equation}\label{eq_a:K}
K
= 
\begin{pmatrix}
0 & -i\beta_{12} \\
-i\beta_{21} & 0
\end{pmatrix}\,.
\end{equation}
This $K$ matrix satisfies the Hermitian condition, i.e. $K=K^{+}$, if and only if:
\begin{equation}\label{eq_a:betaHermitian}
\beta_{12} = -\beta_{21}^{*} =: \beta\,.
\end{equation}
Then, reformulating the backscattering coefficient as $\beta = \left|\beta\right|e^{i\varphi}$ is easy to obtain:
\begin{align}
g_{1} = i \left|\beta\right|\sin{\left[\theta\right]}\quad&\Rightarrow\quad\Real[g_{1}]=0\,,\label{eq_a:gh1}\\
g_{2} = \left|\beta\right|\cos{\left[\theta\right]}\quad&\Rightarrow\quad\Imag[g_{2}]=0\label{eq_a:gh2}\,,
\end{align}
where $\theta = \phi+\varphi = \phi+arg[\beta]$.
Consequently, the eigenvectors changed in:
\begin{equation}\label{eq_a:Eigenvectors_g}
v_{1,2} = \frac{1}{\sqrt{1+\left|\frac{\sin{\left[\theta\right]}\mp1}{\cos{\left[\theta\right]}}\right|^2}}
\begin{pmatrix}
i \frac{\sin{\left[\theta\right]}\mp1}{\cos{\left[\theta\right]}}\\
1
\end{pmatrix}\,.
\end{equation}
On the other hand, the eigenvalues reduce to $\lambda_{1,2} = (\omega_{0}\pm |\beta|)-i(\gamma+\Gamma)$.
As mentioned in the main text, the eigenvectors depend just on the coefficient $\theta$ and are not affected by the strength of the coupling $\beta$. Furthermore, the inner product is equal to zero $\braket{\bm{v}_1|\bm{v}_2}=0$, and therefore, when $\beta=0$ the system works on a diabolic point.  

\section{Experimental procedure} 
\label{app:exp}
In this section, we expose the experimental procedure used to obtain a symmetric interferometric excitation.
First, we measured the losses of each component of the experimental setup and we calibrated the responses of the detectors.
Then we coupled light into the microresonator and, using the detectors ${\rm PD_{C1}}$, ${\rm PD_{C2}}$, ${\rm PD_{L}}$ and ${\rm PD_{R}}$ and VOAs, we set the intensity in the two arms almost equal and low enough not to observe nonlinear effects.
At this point, the first step is to balance the length of the two optical paths after the first fiber splitter. This is needed to decrease the relative phase variation between the two input fields as a function of wavelength $(2\pi n\Delta L/\lambda)$, and thus, to achieve an almost constant relative phase $\phi$ for a microresonator resonant wavelength.
Therefore, we performed repeated wavelength scans, using a scan rate of $100~{\rm nm/s}$, over the entire spectrum and we changed the optical length of the right arm to find the condition where the period of the observed oscillations at the ${\rm PD_{R}}$ and ${\rm PD_{L}}$ detectors is maximum.

Once we set the delay line, we used the following procedure to input the same intensities in the bus waveguide sides, i.e. to perform a symmetric interferometric excitation.
\begin{enumerate}
	\item We block one of the two arms of the setup and then we maximize the transmission of the device at a non-resonant wavelength ($1530\nm$).
	\item We repeat this for the other arm.
	\item We set the VOAs to ensure the same transmissions from both device outputs (taking into account different circulators losses and detectors sensitivities).
	\item We determine the ratio of the powers at the two detectors ${\rm PD_{C1}}$ and ${\rm PD_{C2}}$ ($I_{\rm 3,C1}/I_{\rm 3,C2}$).
	\item We set the VOAs to obtain the same reflections from both device outputs.
	\item We determine the ratio of the powers at the two detectors ${\rm PD_{C1}}$ and ${\rm PD_{C2}}$ ($I_{\rm 5,C1}/I_{\rm 5,C2}$).
	\item By using these data, we e derive the relation between the grating couplers coefficients:
	\begin{equation}
	\frac{g_{\rm L}}{g_{\rm R}}=\sqrt{\frac{I_{\rm 5,C2}}{I_{\rm 5,C1}}\frac{I_{\rm 3,C1}}{I_{\rm 3,C2}}}\,,
	\end{equation}
	where $g_{\rm L}$ and $g_{\rm R}$ are the coupling coefficients for the left and right gratings at the bus waveguide edges. 
	\item Considering the ratio of the coupling coefficients ($g_{\rm L}/g_{\rm R}$), we set the VOAs such that the field amplitudes within the bus waveguide are the same for both the excitation directions.
\end{enumerate}

\section*{Acknowledgement}

We acknowledge funding from Ministero dell\textquotesingle Istruzione, dell\textquotesingle Universit\`a e della Ricerca (PRIN PELM (20177 PSCKT)) and by PAT through the Q@TN joint lab. We gratefully thank Dr. Iacopo Carusotto and Mr.  Alberto Mu$\rm\tilde{n}$oz de las Heras for useful inputs, valuable comments and interesting discussion, as well as Mr. Enrico Moser for technical support.
This work was supported by Q@TN, the joint lab between University of Trento, FBK- Fondazione Bruno Kessler, INFN- National Institute for Nuclear Physics and CNR- National Research Council.

\bibliographystyle{unsrt}
\bibliography{Bibliography.bib}

\end{document}